\documentclass[singlespacing]{elsart}

\usepackage{graphicx}

\usepackage{amssymb}

\begin{document}
\sloppy

\begin{frontmatter}



\title{Analysis of Low-Momentum Correlations with Cartesian
Harmonics}


\author[1,2]{P.\ Danielewicz}
\author[2]{and S.\ Pratt}

\address[1]{National Superconducting Cyclotron Laboratory, Michigan State University,
East Lansing, MI 48824-1321, USA}
\address[2]{Department of Physics and Astronomy, Michigan State University,
East Lansing, MI 48824-2320, USA}

\begin{abstract}
Exploiting final-state interactions and/or identity interference,
analysis of anisotropic correlations of particles at low-relative
velocities yields information on the anisotropy of emission sources in
heavy-ion reactions.  We show that the use of cartesian surface-spherical
harmonics in such analysis allows for a systematic expansion of the
correlations in terms of real angular-moment coefficients dependent on
relative momentum. The coefficients are directly related to the analogous
coefficients for emission sources. We illustrate the analysis with an
example of correlations generated by classical Coulomb interaction.
\end{abstract}

\begin{keyword}
correlation \sep interferometry \sep intensity interferometry \sep cartesian
harmonics \sep anisotropic correlation

\PACS
25.70.Pq \sep 25.75.Gz

\end{keyword}
\end{frontmatter}

Correlations of particles at low relative velocities are commonly used for
assessing geometric features of emission zones in nuclear reactions with
multiparticle final states \cite{Heinz:1999rw,Bauer:1993wq}. At low relative
momentum, the correlations exhibit structures, due to final-state interactions
and/or identity interference, that are more pronounced for tighter emission
zones. When particle emission zones are deformed, corresponding shape
anisotropies are observed in the correlation function which can be studied as a
function of the orientation of the relative momentum. Measured anisotropies in
the correlation functions thus provide insight into the shape of the emission
zone which is intimately connected to such aspects of the emission as reaction
geometry, collective expansion, emission duration and differences in emission
times for different species \cite{lednicky,panitkinoffset,gelderloos}.

Final-state interactions and identity interference link the measured
correlation function, ${\mathcal R}_{\bf P}({\bf q})$ (or alternatively
${\mathcal C}_{\bf P}({\bf q})$), to the source function, ${\mathcal S}_{\bf
P}({\bf r})$, which provides the probability that two particles of the same
velocity, whose total momentum is ${\bf P}$, are separated by a distance ${\bf
r}$ at emission \cite{Koonin:1977fh,Anchishkin:1997tb}:
\begin{equation}
{\mathcal R}_{\bf P}({\bf q})  \equiv  {\mathcal C}_{\bf P}({\bf q}) - 1
\equiv \frac{\frac{{\rm d}^6N^{ab}}{{\rm d}^3p_a \,
{\rm d}^3p_b}}{\frac{{\rm d} N^a}{{\rm d}^3p_a} \,
\frac{{\rm d}N^b}{{\rm d}^3p_b}}-1
 =  \int {\rm d}^3r \,
 \left[|\phi_{\bf q}^{(-)} ({\bf r})|^2-1\right] \,
 {\mathcal S}^{ab}_{\bf P}({\bf r}) \, .
\label{eq:master}
\end{equation}
Here, ${\bf q}$ and ${\bf r}$ are the relative momentum and relative spatial
separation as determined by an observer in the two-particle rest frame and
$\phi^{(-)}$ is the relative wave function for the asymptotic momentum ${\bf
q}$.  For particles with intrinsic spins, the square $|\phi^{(-)}|^2$ is
averaged over spins, ensuring its dependence only on $q$, $r$
and the angle between the vectors, $\theta_{{\bf q} \, {\bf r}}$.

Various means have been employed to analyze anisotropies in correlations to
provide quantitative measurement of the anisotropies in a source function. When
analyzing identical pion correlations, the most common means has been to
present projections of the correlation function along fixed directions of the
relative momentum while constraining the remaining momentum components. Such
projections often neglect regions of the data outside of the employed cuts and
they can be clumsy if one is unaware of the best directions to orient the
coordinate system. A second means for analyzing anisotropies involves fitting
to three-dimensional parameterizations, such as for gaussian or blast-wave
models \cite{Retiere:2003kf}. Since source parameterizations typically employ
on the order of 10 parameters, it can be difficult to determine the confidence
of the fits due to the inherent complexity of cross correlations.

A third class of approaches involves applying tesseral spherical harmonics
$Y_{\ell m}$ (regarding terminology for harmonics see~\cite{applequist02}),
to express the directional information in the correlation
function and in the source function~\cite{Brown:1997ku},
\begin{eqnarray}
{\mathcal R}_{\ell m}(q) &=&
(4\pi)^{-1/2}\int {\rm d}\Omega_{\bf q} \,
Y_{\ell m}^*(\Omega_{\bf q}) \,
{\mathcal R}({\bf q})
\, , \nonumber \\
{\mathcal S}_{\ell m}(r) & = & (4\pi)^{-1/2}\int {\rm d}\Omega_{\bf r} \,
Y_{\ell m}^*(\Omega_{\bf r} ) \,
{\mathcal S}({\bf r}).
\end{eqnarray}
Here, the labels indicating the total momentum ${\bf P}$ and the particle types
are suppressed.  In terms of the projections, Eq.\ (\ref{eq:master}) can be
re-expressed \cite{Brown:1997ku} as
\begin{equation}
\label{eq:masterylm}
{\mathcal R}_{\ell m}(q) = 4 \pi \int  {\rm d} r \, r^2 \,
K_\ell(q,r) \,
{\mathcal S}_{\ell m}(r) \, ,
\end{equation}
where
\begin{equation}
K_\ell(q,r) =  \frac{1}{2} \int {\rm d} \cos\theta_{\bf qr} \,
P_\ell(\cos\theta_{\bf qr}) \,
\left[|\phi^{(-)} (q,r,\cos\theta_{\bf qr})|^2-1\right] \, .
\label{eq:Kl}
\end{equation}
Since a
given $\ell m$ projection of ${\mathcal R}$ is only related to the same $\ell
m$ projection of ${\mathcal S}$, the complexity of deconvoluting the shape
information present in ${\mathcal R}$ is enormously simplified.

The disadvantage of expansion in the tesseral harmonics $Y_{\ell m}$ is that the connection
between the geometric features of the real source function ${\mathcal S}({\bf
r})$ and the complex valued projections ${\mathcal S}_{\ell m}(r)$ is not
transparent.  The $Y_{\ell m}$ harmonics are convenient for
analyzing quantum angular momentum, but are clumsy for expressing anisotropies
of real-valued functions.
 To rectify the shortcoming we propose performing a similar
analysis using real-valued cartesian surface-spherical harmonics
\cite{applequist89,applequist02} that directly address moments of the analyzed anisotropic functions.
Cartesian harmonics are based on the products
of unit vector components, $n_{\alpha_1}\, n_{\alpha_2}\cdots n_{\alpha_\ell}$.
Due to the normalization identity $n_x^2 + n_y^2 + n_z^2=1$, at a given~$\ell
\ge 2$, the different component products are not linearly independent as
functions of spherical angle; at a given $\ell$, the products are spanned by
tesseral harmonics of rank $\ell' \le \ell$, with $\ell'$ of the same evenness
as~$\ell$. Projecting away the lower-rank tesseral components, $\ell'< \ell$,
is a linear operation in the cartesian rank-$\ell$ space, represented
by a rank-$2 \ell$ tensor ${\mathcal P}^{(2\ell)}$ that produces cartesian
rank-$\ell$ harmonics ${\mathcal
A}^{(\ell)}_{\alpha_1\cdots\alpha_\ell}(\Omega)$:
\begin{equation}
\label{eq:projection}
{\mathcal A}_{\alpha_1\cdots \alpha_\ell}^{(\ell)}=
\sum_{\alpha_1' \ldots \alpha_\ell'}
{\mathcal
  P}^{(2\ell)}_{\alpha_1 \ldots \alpha_\ell , \alpha_1'
\ldots\alpha_\ell'} \,
n_{\alpha_1'} \cdots n_{\alpha_\ell'} \, .
\end{equation}
Tesseral harmonics of rank $\ell$, are, conversely, expressible in terms of the
powers of unit-vector components of the order $\ell' \le \ell$, with $\ell'$ of
the same evenness as~$\ell$.  In the end, the harmonic ${\mathcal A}^{(\ell)}$
from (\ref{eq:projection}) may be expressed as a combination of the powers of
$n_\alpha$ of order $\ell' \le \ell$, where $\ell'$ is of the same evenness as
$\ell$, as shown in Table~\ref{table:cartesian}. The leading term is a simple
product of unit vectors with the same indices as ${\mathcal A}$. Since
cartesian harmonics ${\mathcal A}^{(\ell)}$ are linear combinations of tesseral
harmonics of rank $\ell$, the product $r^\ell \, {\mathcal A}^{(\ell)}$
satisfies the Laplace equation.
\begin{table}
\caption{\label{table:cartesian}Cartesian surface-spherical harmonics of rank
$\ell\le 4$.  Other harmonics can be found by permuting indices or switching
indices, i.e., $x\leftrightarrow y$, $x\leftrightarrow z$ or $y\leftrightarrow
z$.}
\begin{tabular}{|c|c|}\hline
${\mathcal A}^{(1)}_x=n_x$
& ${\mathcal A}^{(3)}_{xyz}=n_x \, n_y \, n_z$ \\
${\mathcal A}^{(2)}_{xx}=n_x^2-1/3$
& $A^{(4)}_{xxxx}=n_x^4-(6/7)n_x^2+3/35$ \\
${\mathcal A}^{(2)}_{xy}=n_x \, n_y$
& ${\mathcal A}^{(4)}_{xxxy}=n_x^3 \, n_y-(3/7)n_x \, n_y$\\
$A^{(3)}_{xxx}=n_x^3-(3/5)n_x$
&  ${\mathcal A}^{(4)}_{xxyy}=n_x^2 \, n_y^2-(1/7)n_x^2-(1/7)n_y^2+1/35$\\
${\mathcal A}^{(3)}_{xxy}=n_x^2 \, n_y-(1/5)n_y$
& ${\mathcal A}^{(4)}_{xxyz}=n_x^2 \, n_y \, n_z-(1/7)n_y \, n_z$\\ \hline
\end{tabular}
\end{table}

Because ${\mathcal A}^{(\ell)}$ is a linear combination of the tesseral
harmonics of rank $\ell$, and because the kernel in Eq.\ (\ref{eq:masterylm})
depends only on $\ell$ (and not $m$), an analogous equation to
(\ref{eq:masterylm}) can be derived for cartesian harmonic coefficients:
\begin{equation}
\label{eq:mastercar}
{\mathcal R}^{(\ell)}_{\alpha_1\cdots\alpha_\ell}(q)
= 4 \pi \int  {\rm d}r \, r^2 \,
K_\ell(q,r) \,
{\mathcal S}^{(\ell)}_{\alpha_1\cdots\alpha_\ell}(r) \, ,
\end{equation}
where the coefficients are defined as
\begin{eqnarray}
{\mathcal R}^{(\ell)}_{\alpha_1\ldots\alpha_\ell}(q)
& = &\frac{(2\ell+1)!!}{\ell !}\int \frac{{\rm d}\Omega_{\bf q}}{4\pi} \,
{\mathcal A}^{(\ell)}_{\alpha_1\ldots\alpha_\ell}(\Omega_{\bf q)} \,
{\mathcal R}({\bf q}) \, ,
\nonumber
\\
{\mathcal S}^{(\ell)}_{\alpha_1\ldots\alpha_\ell}(r)
&=&\frac{(2\ell+1)!!}{\ell!}\int \frac{{\rm d}\Omega_{\bf r}}{4\pi} \,
{\mathcal A}^{(\ell)}_{\alpha_1\ldots\alpha_\ell}(\Omega_{\bf r}) \,
{\mathcal S}({\bf r}) \,.
\label{eq:RSdef}
\end{eqnarray}

Cartesian harmonics satisfy the completeness relations:
\begin{eqnarray}
\delta(\Omega'-\Omega)&=&\frac{1}{4\pi}\sum_{\ell}
\frac{(2\ell+1)!!}{\ell !}  \sum_{\alpha_1 \ldots \alpha_\ell}  {\mathcal
A}_{\alpha_1\ldots\alpha_\ell}^{(\ell)}(\Omega') \, {\mathcal
A}_{\alpha_1\ldots\alpha_\ell}^{(\ell)}(\Omega)
\nonumber
\\
&=&\frac{1}{4\pi}\sum_{\ell}
\frac{(2\ell+1)!!}{\ell !} \sum_{\alpha_1 \ldots \alpha_\ell}
{\mathcal A}^{(\ell)}_{\alpha_1\ldots\alpha_\ell}(\Omega') \,
n_{\alpha_1} \cdots n_{\alpha_\ell} \, .
\label{eq:completeness}
\end{eqnarray}
Replacing ${\mathcal A}^{(\ell)}(\Omega)$ with the product
$n_{\alpha_1}\cdots n_{\alpha_\ell}$ was justified by the fact that the
projection operator is employed in ${\mathcal A}^{(\ell)}(\Omega')$. Equation
(\ref{eq:completeness}) allows one to restore ${\mathcal R}({\bf q})$ and
${\mathcal S}({\bf r})$ from the cartesian coefficients,
\begin{eqnarray}
\nonumber
{\mathcal R}({\bf q}) & = &\sum_{\ell} \sum_{\alpha_1\ldots\alpha_\ell}
{\mathcal R}_{\alpha_1\ldots\alpha_\ell}^{(\ell)} \,
\hat{q}_{\alpha_1}\cdots \hat{q}_{\alpha_\ell} \, ,\\
{\mathcal S}({\bf r}) & = &\sum_{\ell} \sum_{\alpha_1\ldots\alpha_\ell}
{\mathcal S}_{\alpha_1\ldots\alpha_\ell}^{(\ell)} \,
\hat{r}_{\alpha_1}\cdots \hat{r}_{\alpha_\ell} \, ,
\label{eq:RSexp}
\end{eqnarray}
where $\hat{\bf q}$ and $\hat{\bf r}$ are unit vectors in the directions of
${\bf q}$ and ${\bf r}$.

Cartesian harmonics have other important properties.  They are symmetric and
traceless, i.e., if one sums ${\mathcal A}^{(\ell)}_{\alpha_1
\cdots\alpha_\ell}$ over any two equated indices, the result is zero. In fact, the
projection operator ${\mathcal P}$ in Eq.\ (\ref{eq:projection}) is also
referred to as a de-tracing operator. Tracelessness ensures, on its own, that
$r^\ell \, {\mathcal A}^{(\ell)}$ satisfies the Laplace equation.  While the
$z$ axis is singled out in the construction of tesseral harmonics, the axes are
treated symmetrically in the construction of cartesian harmonics, and different
components can be obtained from each other by interchanging the $x$, $y$ and
$z$ axes. For instance, the expression for ${\mathcal A}_{yzz}^{(\ell=3)}$ can
be found by replacing replacing $n_x$ with $n_z$ in the expression for
${\mathcal A}_{xxy}^{(\ell=3)}$ in Table \ref{table:cartesian}. Cartesian
harmonics can be also generated recursively, beginning with ${\mathcal
A}^{(\ell=0)}=1$,
\begin{eqnarray}
\nonumber
{\mathcal A}_{\alpha_1\ldots\alpha_\ell}^{(\ell)} (\Omega)&=& \frac{1}{\ell}
\sum_i n_{\alpha_i} \, {\mathcal
A}_{\alpha_1\ldots\alpha_{i-1} \, \alpha_{i+1}\ldots\alpha_\ell}^{(\ell-1)}
(\Omega) - \frac{2}{\ell(2\ell-1)}  \\
&&\times \sum_{i<j}
\sum_{\alpha}\delta_{\alpha_i \, \alpha_j} \,
n_\alpha \,
{\mathcal A}_{\alpha \, \alpha_1\ldots\alpha_{i-1} \, \alpha_{i+1}\ldots
\alpha_{j-1} \, \alpha_{j+1}\ldots\alpha_\ell}^{(\ell-1)}(\Omega) \, .
\end{eqnarray}
The above recursion can be used to obtain the coefficients for expanding the
cartesian harmonics in terms of unit vector components \cite{applequist02}. The
strategy of recursion, carried out in the other direction, leads to the
coefficients of expansion of the products in terms of harmonics:
\begin{eqnarray}
\label{eq:nnnA}
n_x^{\ell_x} \, n_y^{\ell_y} \, n_z^{\ell_z} &=&
\sum_{m_\alpha \le \ell_\alpha/2}
\chi_{m_x \, m_y \, m_z}^{\ell_x \, \ell_y \, \ell_z} \,
{\mathcal A}_{[\ell_x-2m_x,\ell_y-2m_y,\ell_z-2m_z]}^{(\ell-2m)} \, ,\\
\nonumber
\chi_{m_x \, m_y \, m_z}^{\ell_x \, \ell_y \, \ell_z} &\equiv&
\frac{(2 \ell - 4 m + 1)!!}{2^m \, (2 \ell - 2 m + 1)!!} \,
\prod_{\alpha=x,y,z} \frac{\ell_\alpha !}{m_\alpha!
\, (\ell_\alpha - 2 m_\alpha)!} \, .
\end{eqnarray}
Here, $(-1)!!=1$, $m=m_x+m_y+m_z$, $\ell=\ell_x+\ell_y+\ell_z$ and
$[\ell_x,\ell_y,\ell_z]$ is any set of indices where $x$ occurs $\ell_x$ times,
$y$ occurs $\ell_y$ times and $z$ occurs $\ell_z$ times.  

The expansion (\ref{eq:nnnA}) is
useful for calculating moments of ${\mathcal S}({\bf r})$,
\begin{eqnarray}
\nonumber
\int {\rm d}^3r \,  x^{\ell_x} \, y^{\ell_y} \, z^{\ell_z} \,
{\mathcal S}({\bf r}) \left/
\int {\rm d}^3r \, {\mathcal S}({\bf r}) \right.
&=& \sum_{m_\alpha \le \ell_\alpha/2} \frac{(\ell - 2m)!}{(2\ell-4m+1)!!} \,
\chi_{m_x,m_y,m_z}^{\ell_x,\ell_y,\ell_z}  \\
&& \hspace*{-140pt}
\times \int {\rm d}r \, r^{\ell+2} \,
{\mathcal S}_{[\ell_x-2m_x,\ell_y-2m_y,\ell_z-2m_z]}^{(\ell-2m)} (r)
\left/ \int {\rm d}r \, r^{2} \, {\mathcal S}^{(0)}(r) \right.
\label{eq:moments} \, .
\end{eqnarray}
In expansion in terms of cartesian harmonics, reflection symmetries with
respect to the coordinate axes manifest themselves explicitly.  Thus, 
if e.g.\ the source function is symmetric under the $y \rightarrow -y$ transformation,
all odd-$\ell_y$ coefficients of ${\mathcal S}$ vanish.

Since rank-$\ell$ cartesian harmonics are linear combinations of rank-$\ell$
tesseral harmonics, cartesian harmonics of different $\ell$ are orthogonal to
one another. Within a given rank, though, there are $(\ell+1)(\ell+2)/2$
combinations of $[\ell_x,\ell_y,\ell_z]$, while only $(2 \ell +1)$ components
can be independent. As an independent set, one can choose the components
characterized by $\ell_\alpha=0,1$, for a selected coordinate $\alpha$, and the
remaining harmonics within the rank can be generated via the traceleness
property. For example, given harmonics for $\ell_x=0,1$, the application of ${\mathcal
A}_{[\ell_x+2,\ell_y,\ell_z]}=-{\mathcal
A}_{[\ell_x,\ell_y+2,\ell_z]}-{\mathcal A}_{[\ell_x,\ell_y,\ell_z+2]}$ yields
the other harmonics. As for scalar product values within a rank, they can be
found in \cite{applequist02}.

For weak physical anisotropies it is likely that only the few lowest terms of
angular expansion can be ascribed to the correlation functions or the source.
For $\ell \le 2$, the results can be summarized in terms of amplitudes and
direction vectors as a function of $q$ and $r$, as illustrated here with the
case of ${\mathcal R}$:
\begin{equation}
{\mathcal R}({\bf q})  =
{\mathcal R}^{(0)}(q) + \sum_\alpha \, {\mathcal R}^{(1)}_\alpha(q) \,
\hat{q}_\alpha + \sum_{\alpha_1 \, \alpha_2}
{\mathcal R}^{(2)}_{\alpha_1 \alpha_2} (q) \,
\hat{q}_{\alpha_1} \, \hat{q}_{\alpha_2} + \ldots \, .
\end{equation}
The function ${\mathcal R}^{(0)}$ is the correlation averaged over angles. The
${\mathcal R}^{(1)}$ components describe the dipole anisotropy and can be
summarized in terms of the direction ${\bf e}^{(1)}(q)$ and magnitude
$R^{(1)}(q)$, ${\mathcal R}^{(1)}_\alpha(q) = R^{(1)} \, e^{(1)}_\alpha$.  The
${\mathcal R}^{(2)}$ components describe the quadrupole anisotropy and can be
summarized in terms of three orthogonal principal direction vectors $\lbrace
{\bf e}_k^{(2)} (q) \rbrace_{k=1,2,3}$ and three distortions that sum up to
zero due to tracelessness: ${\mathcal R}^{(2)}_{\alpha \beta} (q) =R_1^{(2)} \,
e_{1 \alpha}^{(2)} \, e_{1 \beta}^{(2)} + R_3^{(2)} \, e_{3 \alpha}^{(2)} \,
e_{3 \beta}^{(2)} - \left( R_1^{(2)} + R_3^{(2)} \right) \, e_{2 \alpha}^{(2)}
\, e_{2 \beta}^{(2)} $.  For higher-order distortions, the direction of maximal
distortion ${\bf e}_3^{(\ell)}$ may be defined as one maximizing the $\ell$'th
cartesian harmonic projected in the direction of ${\bf e}_3$: $ \sum_{\alpha_1
\ldots \alpha_\ell} {\mathcal R}_{\alpha_1 \ldots \alpha_\ell}^{(\ell)} \, {\bf
e}_{3 \alpha_1}^{(\ell)} \cdots {\bf e}_{3 \alpha_\ell}^{(\ell)}$.  The
condition of maximum implies $\sum_{\alpha_1 \ldots \alpha_\ell} {\mathcal
R}_{\alpha_1 \, \alpha_2 \ldots \alpha_\ell}^{(\ell)} \, e_{k \,
\alpha_1}^{(\ell)} \, e_{3 \, \alpha_2}^{(\ell)} \cdots e_{3 \,
\alpha_n}^{(\ell)}$ $= 0$, where $k=1,2$ and ${\bf e}_1^{(\ell)}$ and ${\bf
e}_2^{(\ell)}$ span the space of directions perpendicular to ${\bf
e}_3^{(\ell)}$.  The direction of ${\bf e}_1^{(\ell)}$ may be chosen to
minimize the projection of ${\mathcal R}^{(\ell)}$ in the plane perpendicular
to ${\bf e}_3^{(\ell)}$.

\begin{figure}
\begin{center}
\protect\includegraphics[width=.5\linewidth]{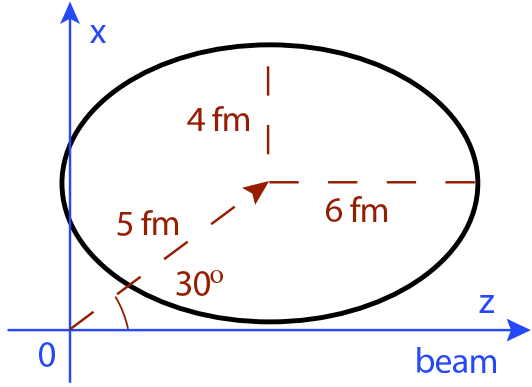}
\end{center}
\caption{Characteristics of the sample source.  The ellipse represents the
contour of constant density within $x$-$z$ plane, at half the maximum density.}
\label{fig:sourcech}
\end{figure}
\begin{figure}
\begin{center}
\protect\includegraphics[width=.5\linewidth]{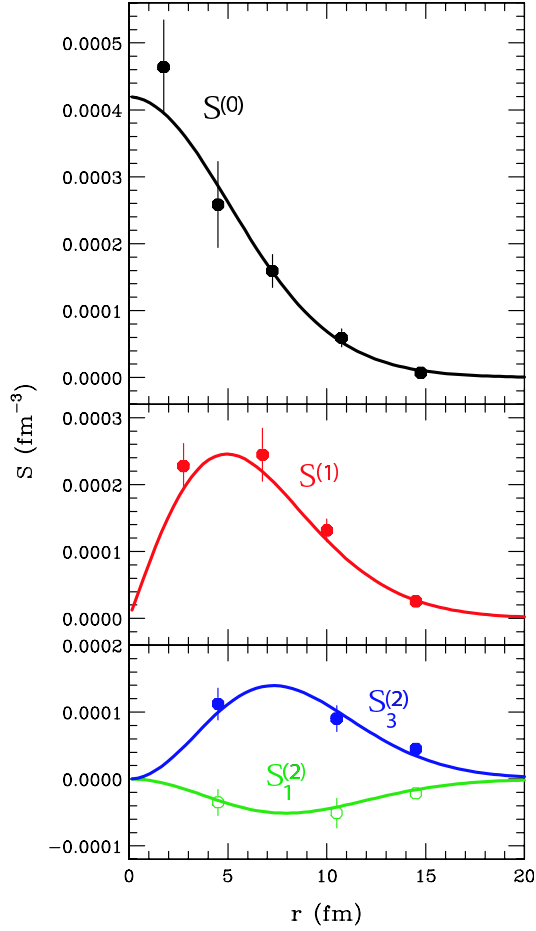}
\end{center}
\caption{Low-$\ell$ values of the sample relative source, as a function of the
relative distance $r$.  Lines and symbols represent, respectively, exact values
and values obtained through imaging of an anisotropic correlation function.
Top panel shows the source averaged over angles, $S^{(0)}$.  Middle panel shows
the dipole distortion $S^{(1)}$.  Bottom panel shows the larger, $S_3^{(2)}$,
and smaller, $S_1^{(2)}$, quadrupole distortions within the $x$-$z$ plane.  }
\label{fig:sourcemo}
\end{figure}
\begin{figure}
\begin{center}
\protect\includegraphics[width=.5\linewidth]{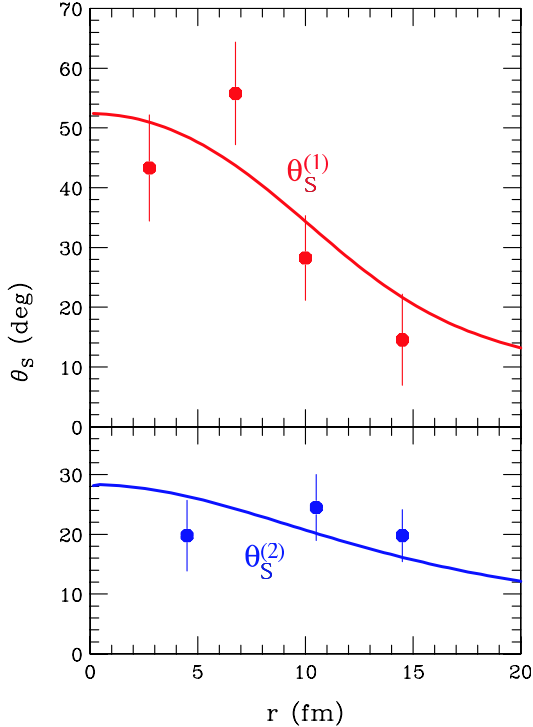}
\end{center}
\caption{Angles characterizing the dipole (top) and quadrupole (bottom)
distortions of the sample relative source, as a function of the relative
distance $r$.  Lines and symbols represent, respectively, exact angles and
angles obtained from imaging of an anisotropic correlation function.  }
\label{fig:sourcean}
\end{figure}

We next illustrate with an example how the correlation functions and source can
be quantified following the cartesian coefficients.
We choose a coordinate system with the $z$-axis along the beam direction and
$x$-axis along the pair transverse momentum ${\bf P}^\perp$.  We take a sample
relative source for the particles, in a gaussian form with a larger 6 fm width
in the beam direction and lower 4 fm widths in the transverse directions, cf.\
Fig.\ \ref{fig:sourcech}.  The source is displaced from the origin by 5 fm,
along ${\bf P}$ in system cm, assumed to point at 30$^\circ$ relative to the
beam axis.  The cartesian coefficients are calculated for the source
according to Eq.\ (\ref{eq:RSexp}) and expressed in terms of amplitudes and
angles shown in Figs.\ \ref{fig:sourcemo} and~\ref{fig:sourcean}.  The displayed dipole
characteristics result then from:  ${S}^{(1)} \,
\sin{\theta_S^{(1)}}= {\mathcal S}_x^1 = 3 \int \frac{{\rm d}\Omega}{4\pi} \,
{\mathcal S}({\bf r}) \sin{\theta}$ and $S^{(1)} \, \cos{\theta_S^{(1)}}=
{\mathcal S}_z^{(1)} = 3 \int \frac{{\rm d}\Omega}{4\pi} \, {\mathcal S}({\bf
r}) \cos{\theta}$, and quadrupole from: ${\mathcal S}_{xx}^{(2)} = S_1^{(2)}
\cos^2{\theta_S^{(2)}} + S_3^{(2)} \sin^2{\theta_S^{(2)}}$, ${\mathcal
S}_{zz}^{(2)} = S_1^{(2)} \sin^2{\theta_S^{(2)}} + S_3^{(2)}
\cos^2{\theta_S^{(2)}}$ and ${\mathcal S}_{xz}^{(2)} = \left( S_3^{(2)} -
S_1^{(2)} \right) \sin^2{\theta_S^{(2)}} $, where ${\mathcal S}_{xx}^{(2)} =
\frac{15}{2} \int \frac{{\rm d}\Omega}{4\pi} \, {\mathcal S}({\bf r}) \left(
\sin^2 {\theta} \cos^2 {\phi} - \frac{1}{3} \right)$, ${\mathcal S}_{zz}^{(2)}
= \frac{15}{2} \int \frac{{\rm d}\Omega}{4\pi} \, {\mathcal S}({\bf r}) \left(
\cos^2 {\theta} - \frac{1}{3} \right)$ and ${\mathcal S}_{xz}^{(2)} =
\frac{15}{2} \int \frac{{\rm d}\Omega}{4\pi} \, {\mathcal S}({\bf r})
\cos{\theta} \sin {\theta} \cos {\phi} $.

Despite the fact that the maximum of source density ${\mathcal S}$ occurs away from
${\bf r}=0$, the angle-averaged source $S^{(0)}$
in Fig.~\ref{fig:sourcemo} is maximal at $r=0$.
An analytic source function ${\mathcal S}$ may be Taylor
expanded in three dimensions about ${\bf r}=0$.
The lowest-order terms of the Taylor expansion that can contribute to a rank-$\ell$
cartesian coefficient of ${\mathcal S}$ must involve an $\ell$'th order
derivative and rise as~$r^\ell$,
as is apparent in Fig.~\ref{fig:sourcemo}.  The subsequent contributing terms
rise as $r^{\ell +2k}$, $k=1,2,\ldots$, which is important if one tries to parameterize
the functions~$S^{(\ell)}$.

At low $r$, the angles $\theta_S^{(1)}$ and $\theta_S^{(2)}$ are determined
by the derivatives of ${\mathcal S}$ at ${\bf r}=0$; in particular $\theta_S^{(1)}$
gives the direction of the gradient.
At high $r$, the angles follow the elongation of the gaussian
and, correspondingly,
they approach zero as $r\rightarrow\infty$.

Next, we determine correlation function coefficients for our sample source,
examine how the coefficients reflect the source features and attempt to restore
those features through imaging. We choose the classical limit of repulsive Coulomb
interactions for the emitted particles, such as appropriate for intermediate
mass fragments \cite{kim,Pratt:2003ar}.  The kernel for Eqs.\ (\ref{eq:master})
and (\ref{eq:Kl}) can be analytically calculated by considering changes in momentum
volume associated with Coulomb trajectories, $|\phi|^2 = {\rm d^3}\, q_0/
{\rm d^3}\, q$, where ${\bf q}_0$ is the starting momentum at the separation ${\bf r}$.
The result depends only on the angle $\theta_{{\bf q} {\bf r}}$ and
on the separation at emission scaled with the distance of closest approach in a
head-on collision, $r/r_c$, where $q^2/2m_{ab}=Z_a \, Z_b \, e^2/4\pi
\epsilon_0 \, r_c$:
\begin{equation}
\label{eq:KCoul}
|\phi^{(-)}(q,r,\cos\theta_{\bf qr})|^2=
\frac{\Theta \left( 1 + \cos{\theta_{{\bf q} {\bf r}}}-{2 r_c}/{r} \right) \,
(1 + \cos{\theta_{{\bf q} {\bf r}}}-{r_c}/{r})}
{\sqrt{\left(1 + \cos{\theta_{{\bf q} {\bf r}}}\right)^2-
\left(1 + \cos{\theta_{{\bf q} {\bf r}}}\right)
{2 r_c}/{r} } } \, .
\end{equation}
The $\ell=0$ kernel is $K_0 = \Theta(r -r_c) \, \sqrt{1-r_c/r} - 1$, while the
$K_{\ell \ge 1}$ components are, generally, calculated numerically from (\ref{eq:Kl}).

\begin{figure}
\begin{center}
\protect\includegraphics[width=.5\linewidth]{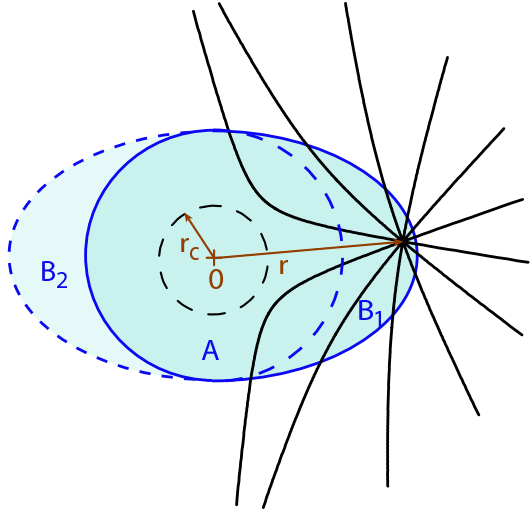}
\end{center}
\caption{A schematic relative source with repulsive Coulomb trajectories
superimposed coming out isotropically from position ${\bf r}$ within the
source.  The region of radius $r_c$ is inaccessible to the trajectories.  The
source consists of an isotropic part~{\sc A}, axially symmetric part {\sc B}$_1$,
responsible for source deformation, and, optionally, part~{\sc B}$_2$ that is a
mirror reflection of {\sc B}$_1$.  }
\label{fig:traj}
\end{figure}
The Coulomb trajectories are illustrated in Fig.~\ref{fig:traj}.
The repulsive Coulomb force principally focuses the trajectories towards
the direction of ${\bf r}$ they originate from.
For $r_c \ll r$ (large $q$),
the $K_\ell$ kernels from from (\ref{eq:Kl}) are affected by the
deflection of trajectories away from
$\cos{\theta_{ \bf qr}} \sim -1$.  With $P_\ell(\cos {\theta} \rightarrow
-1)=(-1)^\ell$, the kernels in this limit are $K_\ell \simeq (-1)^\ell \, K_0
\simeq (-1)^{\ell+1} \, r_c/2r$, with the additional negative sign representing
trajectory depletion.  For $r - r_c \ll r_c$ (low~$q$),
the kernels
are affected by the trajectories bunching up around
$\cos{\theta_{ \bf qr}} \sim 1$.  With $P_\ell(\cos {\theta} \rightarrow
1)= + 1$, the kernels in this limit are $K_{\ell \ge 1} \simeq K_0 + 1 =
\Theta (r - r_c) \, \sqrt{r/r_c - 1}$.  The $K_1$ kernel is always positive and,
thus, the dipole distortion
of the correlation generally points in the same direction as the distortion of the
source, as for e.g.\ the source out of {\sc A} and {\sc B}$_1$ in Fig.~\ref{fig:traj}.
However, the $K_2$
kernel switches sign at $r/r_c=1.63$.  At $r/r_c < 1.63$, a prolate source, such as out
of {\sc A} and {\sc B}$_1$ (and possibly {\sc B}$_2$) in Fig.~\ref{fig:traj}, results
in a prolate correlation function elongated in the same direction.  However, at
$r/r_c > 1.63$, the prolate source results in an oblate correlation with distortion pointing
in transverse directions.

\begin{figure}
\begin{center}
\protect\includegraphics[width=.5\linewidth]{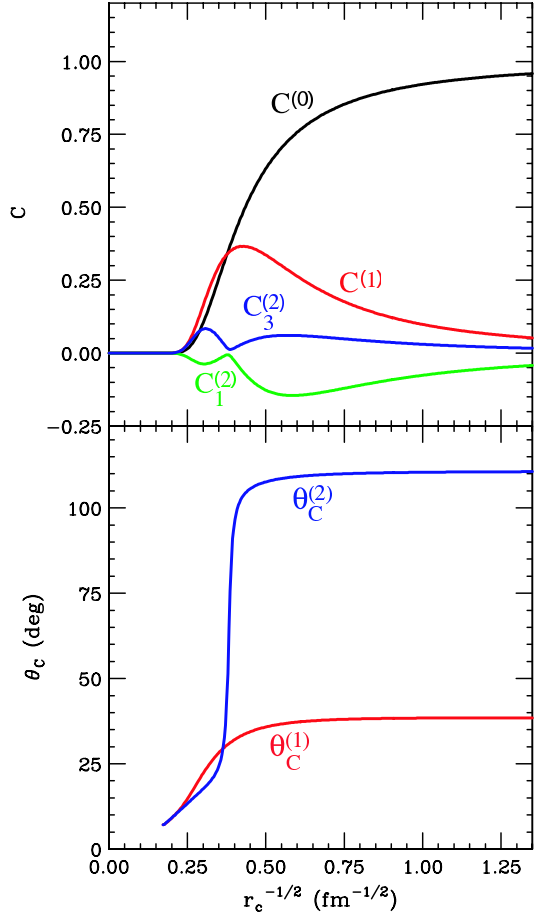}
\end{center}
\caption{Low-$\ell$ characteristics of the Coulomb correlation function for our
sample source: amplitudes (top panel) and angles (bottom panel) as a function
of inverse square root of the head-on return radius $r_c$.  }
\label{fig:corr}
\end{figure}

Figure \ref{fig:corr} shows the characteristics of the correlation function
${\mathcal C}$, in terms of $\ell \le 2$ cartesian coefficients, for the source
of Fig.~\ref{fig:sourcech}, as function of $r_c^{-1/2} \propto q$.
The dependence $\theta_C^{(1)}(r_c)$ in Fig.~\ref{fig:corr} generally retraces
the dependence $\theta_S^{(1)}(r)$ of Fig.~\ref{fig:sourcean}, where $r_c
\lesssim r$, as anticipated.  However, while $\theta_C^{(2)}(r_c)$
starts out in a similar
fashion, it jumps by $90^\circ$ as $K_2$ switches sign, with $r$ in $r/r_c \sim 1.6$
characterizing
the source quadrupole distortion.  The jump is accompanied by the cross-over behavior
for the distortions $C^{(2)}$, expected for the kernel sign change.  In the low and high $q$
limits, the correlation coefficients approach universal integrals of source
coefficients, such as ${\mathcal R}_{\alpha_1 \ldots \alpha_\ell}^{(\ell)}(r_c) =
{\mathcal C}^{ (\ell)}_{\alpha_1 \ldots \alpha_\ell}(r_c) - \delta_{\ell 0}
\simeq 2 \pi r_c \, (-1)^{\ell +1} \int_0^\infty {\rm d}r \, r \, {\mathcal
S}^{ \ell}_{\alpha_1 \ldots \alpha_\ell} (r) $, for $r_c \ll r$.
In that limit, from $r_c^{-1/2} \sim 0.9\,
\mbox{fm}^{-1/2}$ on in Fig.~\ref{fig:corr}, the
relative
magnitudes of $(-1)^\ell \, R^{(\ell)}$ reflect the relative magnitudes of
the $S^{(\ell)}$ distortions for the integrated
traceless tensors.

We next illustrate imaging of source components from the correlation
\cite{Brown:1997ku,Brown:2000aj,Panitkin:2001qb,Verde:2001md,Verde03,chu03}.
The tensorial decomposition allows to use different source
representations within subspaces of different tensorial rank.  With a source
coefficient represented in the basis $\lbrace B_i \rbrace$ as ${\mathcal S}(r)=
\sum_i {\mathcal S}_i \, B_i(r)$, the ${\mathcal R}$ function at $j$'th
discrete value of $q$ (or $r_c^{-1/2}$ in our case), is ${\mathcal R}_j =
K_{ji} \, {\mathcal S}_i$.  Here, following (\ref{eq:mastercar}),
the matrix elements are $K_{ji}= 4 \pi \int_0^\infty {\rm d}r \, r^2
\, K(q_j,r) \, B_i(r)$ and the spherical-tensor
indices are suppressed.  Minimization of $\chi^2 = \sum_j ({\mathcal R}_j -
{\mathcal R}_j^{exp})^2/\sigma_j^2$ with respect to~$\lbrace S_i \rbrace$
yields the result, in matrix form,
\begin{equation}
{\mathcal S}=(K^\top \, \sigma_{\mathcal C}^{-2} \, K)^{-1} \,
K^\top \, \sigma_{\mathcal C}^{-2} \, {\mathcal R} \, ,
\label{eq:SKR}
\end{equation}
where $(\sigma_{\mathcal C}^{-2})_{jk} = \delta_{jk}/\sigma_j^2$.
Equation (\ref{eq:SKR}) is the basis of imaging, i.e.\ determining
${\mathcal S}^{(\ell)}_{\alpha_1\cdots\alpha_\ell}$ values from
${\mathcal R}^{(\ell)}_{\alpha_1\cdots\alpha_\ell}$.
For
illustration, we assume that the $\ell \le 2$ cartesian coefficients of the
correlation function for our sample source have been measured at 80 values of
$r_c^{-1/2}$, between 0 and $3 \, \mbox{fm}^{-1/2}$, subject to an r.m.s.\
error of $\sigma = 0.015$.  For the basis $\lbrace B_i \rbrace$ we take take simple
rectangle functions that, within basis splines, we generally find to provide the most
faithful source values.  As relative errors decrease with increasing
multipolarity, we reduce the number of functions in restoration, from 5 for
$\ell=0$ to 3 for $\ell=2$, all spanning the region $r < 17 \, \mbox{fm}$.
Upon sampling the cartesian coefficients of ${\mathcal R}$, we restore the
coefficients of ${\mathcal S}$ following (\ref{eq:SKR}) and obtain the results
represented by symbols in Figs.\ \ref{fig:sourcemo} and~\ref{fig:sourcean} for
central arguments of the basis functions.  As is apparent, under realistic
circumstances, angular features of the source can be quantitatively restored.

Having the imaged $\ell \le 2$ source coefficients, we can calculate the $\ell
\le 2$ cartesian moments for the imaged region from (\ref{eq:moments}).  For
the image we find, with statistical errors of restoration: $4 \pi \int {\rm d}r
\, r^2 \, {\mathcal S}^{(0)} = 0.99 \pm 0.05$, $\langle x \rangle = (1/3) \int
{\rm d}r \, r^3 \, {\mathcal S}_x^{(1)}/ \int {\rm d}r \, r^2 \, {\mathcal
S}^{(0)} = 2.47 \pm 0.11 \, \mbox{fm}$, $\langle z \rangle = 4.25 \pm 0.13 \,
\mbox{fm}$, $\langle (x - \langle x \rangle)^2 \rangle^{1/2} = 3.80 \pm 0.24 \,
\mbox{fm}$, $\langle y^2 \rangle^{1/2} = 3.81 \pm 0.22 \, \mbox{fm}$, $\langle
(z - \langle z \rangle)^2 \rangle^{1/2} = 5.54 \pm 0.19 \, \mbox{fm}$ and
$\langle (x - \langle x \rangle )( z - \langle z \rangle ) \rangle = 2.23 \pm
1.49 \, \mbox{fm}^2$, which can be compared to the $r<17 \, \mbox{fm}$ results
from the original source of, respectively: 1.00, 2.45~fm, 3.90~fm, 3.99~fm,
4.00~fm, 5.60~fm and $-0.41 \, \mbox{fm}^2$.

In summary, we have discussed the utility of cartesian surface-spherical
harmonics in the analysis of particle correlations at low relative-velocities.
The cartesian harmonics allow for a systematic quantification of anisotropic
correlation functions, through expansion coefficients related to analogous
expansion coefficients for anisotropic emission sources.  For illustrating the
relation, we have employed correlations produced by classical Coulomb
interactions.  To an extent, the features of source anisotropies may be read
off directly from the correlation anisotropies; otherwise, they can be imaged.

\section*{Acknowledgements}

The authors thank David Brown for discussions and for collaboration on a
related project.  This work was supported by the U.S.\ National Science
Foundation under Grant PHY-0245009 and by the U.S.\ Department of Energy under
Grant No. DE-FG02-03ER41259.



\end{document}